\documentclass[a4paper,fleqn]{cas-dc}

\usepackage[numbers]{natbib}

\usepackage{amsmath,amssymb,amsfonts}
\usepackage{algorithmic}
\usepackage{graphicx}
\usepackage{textcomp}
\usepackage{xcolor}
\usepackage{subcaption}  
\usepackage{float}       
\usepackage{algorithm}

\usepackage{booktabs} 

\usepackage{booktabs} 
\usepackage{pgfplots}
\pgfplotsset{compat=1.18} 
\usepackage{pgfplotstable} 
\usepackage{enumitem}

\usepackage{makecell}
\usepackage{footmisc}
\usepackage{hyperref} 

\def\tsc#1{\csdef{#1}{\textsc{\lowercase{#1}}\xspace}}
\tsc{WGM}
\tsc{QE}
\tsc{EP}
\tsc{PMS}
\tsc{BEC}
\tsc{DE}

\begin{document}
\let\WriteBookmarks\relax
\def\floatpagepagefraction{1}
\def\textpagefraction{.001}

\shorttitle{Hybrid AI \& Metaheuristic for Frequency Allocation in 6G UC-CFmMIMO}

\shortauthors{Selina Cheggour et~al.}

\title [mode = title]{Frequency Resource Management in 6G User-Centric CFmMIMO: A Hybrid Reinforcement Learning and Metaheuristic Approach}                      



%

\author[1]{Selina Cheggour}[
                        orcid=0009-0000-6162-1028
                        ]
\ead{selina.cheggour@inria.fr}
\cormark[1]

\author[1]{Valeria Loscri}[]




\affiliation[1]{organization={Inria Lille Nord Europe},
    addressline={Parc scientifique de la Haute-Borne, 40 Av. Halley Bât A}, 
    city={Villeneuve-d'Ascq},
    postcode={59650}, 
    country={France}}

\begin{abstract}
As sixth-generation (6G) networks continue to evolve, AI-driven solutions are playing a crucial role in enabling more efficient and adaptive resource management in wireless communication. One of the key innovations in 6G is user-centric cell-free massive Multiple-Input Multiple-Output (UC-CFmMIMO), a paradigm that eliminates traditional cell boundaries and enhances network performance by dynamically assigning access points (APs) to users. This approach is particularly well-suited for vehicular networks, offering seamless, homogeneous, ultra-reliable, and low-latency connectivity. However, in dense networks, a key challenge lies in efficiently allocating frequency resources within a limited shared subband spectrum while accounting for frequency selectivity and the dependency of signal propagation on bandwidth. These factors make resource allocation increasingly complex, especially in dynamic environments where maintaining Quality of Service (QoS) is critical. This paper tackles these challenges by proposing a hybrid multi-user allocation strategy that integrates reinforcement learning (RL) and metaheuristic optimization to enhance spectral efficiency (SE), ensure fairness, and mitigate interference within shared subbands. To assess its effectiveness, we compare this hybrid approach with two other methods: the bio-inspired Aquila Optimizer (AO) and Deep Deterministic Policy Gradient (DDPG)-based Actor-Critic Reinforcement Learning (AC-RL). Our evaluation is grounded in real-world patterns and channel characteristics, utilizing the 3GPP-3D channel modeling framework (QuaDRiGa) to capture realistic propagation conditions. The results demonstrate that the proposed hybrid strategy achieves a superior balance among competing objectives, underscoring the role of AI-driven resource allocation in advancing UC-CFmMIMO systems for next-generation wireless networks.
\end{abstract}





\maketitle

\section{Introduction}
\sloppy
The advent of sixth-generation (6G) networks marks a significant paradigm shift in wireless communication, characterized by the integration of AI-native solutions and the promise of ultra-reliable, low-latency, and high-capacity connectivity \cite{AINative}. Among the transformative technologies driving this evolution, cell-free massive Multiple-Input Multiple-Output (CFmMIMO) systems stand out for their ability to transcend traditional cell-based network architectures. By deploying spatially distributed antennas that collaboratively serve user equipment (UEs), CFmMIMO eliminates fixed cell boundaries, ensuring seamless connectivity and robust performance in highly dynamic environments \cite{CF1}. However, despite these advantages, CFmMIMO faces a key scalability challenge as the number of UEs and access points (APs) increases. In conventional CFmMIMO, each UE is served by all available APs, leading to excessive computational complexity, increased backhaul signaling, and inefficient resource allocation in dense deployments \cite{CF_Scalability}. As a result, the practicality of CFmMIMO in large-scale networks, such as vehicular environments, becomes limited. To address this limitation, user-centric cell-free massive MIMO (UC-CFmMIMO) has emerged as an enhanced solution. Unlike conventional CFmMIMO, UC-CFmMIMO dynamically assigns each UE to a subset of APs, reducing backhaul overhead and computational complexity while preserving the benefits of distributed Multiple-Input Multiple-Output (MIMO) processing \cite{equalpower1}. This selective association not only improves spectral efficiency (SE) and coverage but also enhances adaptability in highly dynamic environments. This technology is particularly critical for vehicular networks, where uninterrupted communication is essential to support applications like Intelligent Transportation Systems (ITS), autonomous driving, and vehicle-to-everything (V2X) communication.


Despite its potential, UC-CFmMIMO faces several challenges, particularly in the context of resource allocation \cite{CF3}. Vehicular networks are inherently dynamic, characterized by high-speed mobility, rapidly changing UE locations, and fluctuating channel conditions. These factors complicate frequency resource management, leading to challenges in mitigating interference, ensuring fairness, and maximizing SE. Moreover, the scalability of traditional resource allocation methods becomes a pressing issue as the density of UEs and the demand for data continue to increase \cite{CFM}.

Metaheuristic optimization algorithms have long been explored as efficient solutions for resource allocation in complex systems. Techniques such as Genetic Algorithms (GAs) \cite{GAgu} and Ant Colony Optimization (ACO) \cite{ACOou} have been successfully applied to multi-objective problems, demonstrating their ability to balance competing demands such as power control, SE, and interference mitigation. The recently introduced Aquila Optimizer (AO), inspired by the hunting strategies of eagles, offers a promising alternative due to its superior convergence speed and adaptability. AO has been applied to various optimization tasks in wireless networks, including beamforming \cite{AObf} and massive MIMO (mMIMO) systems \cite{AOmm}, where it has shown significant improvements in efficiency and solution quality. \\

In parallel, Reinforcement Learning (RL) has emerged as a transformative approach for addressing resource allocation challenges in next-generation networks. Unlike optimization-based methods, RL operates in a model-free manner, allowing systems to dynamically learn optimal strategies through trial-and-error interactions with the environment \cite{RLra}. This adaptability makes RL particularly well-suited for dynamic and unpredictable scenarios such as vehicular networks. Actor-Critic (AC)-RL models, which combine value-based and policy-based learning, represent a particularly powerful RL paradigm. By balancing exploration (discovering new strategies) and exploitation (optimizing known strategies), AC models have demonstrated effectiveness in optimizing AP-UE association \cite{AC_RL_AP} and power allocation \cite{AC_PO} in CFmMIMO networks.

While both metaheuristic algorithms and RL-based approaches have their strengths, each also has limitations. Metaheuristics like AO excel in rapid convergence and global optimization but may struggle to adapt to dynamic changes in real time. Conversely, RL models are highly adaptable but require extensive training and exploration to achieve optimal performance. Recognizing these complementary strengths, hybrid solutions that integrate metaheuristic algorithms with RL are gaining attention as a promising direction \cite{hybRLmeta}. By leveraging the global optimization capabilities of algorithms like AO and the adaptability of AC-RL, hybrid methods have the potential to address the multi-objective resource allocation problem in UC-CFmMIMO systems more effectively.

\subsection{Related Works}
Recent research in next-generation wireless networks has explored CFmMIMO, RL-based resource allocation, and metaheuristic optimization techniques. Despite these advances, several limitations persist, including the absence of integrated UC-CFmMIMO frameworks, reliance on overly simplistic channel models that overlook 3GPP-3D modeling as well as frequency-dependent channel characteristics, and the lack of hybrid AI-optimization approaches. For example, while the Deep Deterministic Policy Gradient (DDPG) algorithm \cite{DDPGRL} has been effective for power resource allocation in continuous action spaces, its application to UC-CFmMIMO in dense network scenarios remains underexplored and often employs unrealistic channel assumptions. Similarly, works such as \cite{AObf} and \cite{AOmm} have advanced metaheuristic strategies like the Aquila Optimizer for diagonal loading and hybrid beamforming, yet they do not extend to multi-user UC-CFmMIMO environments nor incorporate advanced 3D channel models essential for realistic performance evaluations. Other studies, including \cite{AC_RL_AP}, have applied Soft AC-RL to optimize AP-UE associations and caching in CFmMIMO, but they overlook the frequency-dependent effects of the channels, which are critical in vehicular networks where high mobility introduces significant Doppler shifts \cite{doppler}. In addition, research addressing frequency-selective fading in CFmMIMO-OFDM \cite{FSOFDM} and RL-based resource allocation in V2X networks \cite{CVN, RLV2X}, along with investigations into Next-Generation Multiple Access (NGMA) techniques \cite{NG}, generally fail to consider the densification and expanded bandwidth requirements of modern networks as defined by current numerologies \cite{numerology}. To overcome these challenges, this paper introduces a hybrid RL-metaheuristic strategy that integrates frequency-aware optimization, AI-driven decision-making, and realistic 3GPP-3D channel modeling, thereby providing a scalable and efficient solution for UC-CFmMIMO deployments.

\subsection{Contribution}
This paper builds on these insights by systematically exploring and comparing the performance of the AO, DDPG-based AC-RL, and a novel hybrid approach in the context of UC-CFmMIMO frequency resource allocation. Using realistic channel conditions, we evaluate the scalability, robustness, and efficiency of each approach, providing a comprehensive analysis of their suitability for challenging vehicular networks. The contributions of this work can be summarized as follows:

\begin{itemize}
   \item We introduce a novel UC-CFmMIMO architecture that integrates frequency-selective channel modeling and bandwidth-sharing strategies, specifically tailored for vehicular (V2X) networks. This design ensures a more adaptive spectrum allocation at the physical layer, enabling multiple UEs to efficiently share the same subband while meeting Quality of Service (QoS) constraints. To enhance the realism of our model, we incorporate an advanced 3GPP-3D channel characterization using QuaDRiGa (QUAsi Deterministic RadIo channel GenerAtor), which enables the generation of realistic radio channel impulse responses by capturing the frequency dependency of channel gain, ensuring more precise performance evaluations in vehicular environments.

    \item We formulate a multi-objective optimization problem aimed at maximizing spectral SE, ensuring fairness, and minimizing interference, thereby enhancing resource utilization and improving the UE experience in vehicular network environments.

    \item To solve this optimization problem, we propose a hybrid approach that combines AO and DDPG-based AC-RL. AO is selected for its exploratory capabilities, while AC-RL is chosen for its adaptive learning strengths in handling complex multi-objective problems. Our hybrid strategy is benchmarked against standalone AO and AC-RL, demonstrating significant gains in SE and convergence speed.
\end{itemize}

The rest of this paper is organized as follows: Section \ref{sec:systemDescription} describes the system model, and Section \ref{sec:problemFormulation} presents the problem formulation. Section \ref{sec:adaptiveSolutions} details the proposed algorithms, Section \ref{sec:simulationSetup} outlines the simulation setup, Section \ref{sec:RL_implementation} discusses the implementation of reinforcement learning models, Section \ref{sec:results} presents the results, and Section \ref{sec:conclusion} concludes the paper.

\begin{table}[h!]
    \centering
    \caption{Contrasting Our Contributions to the Literature.}
    \label{tab:related_works}
    \resizebox{\columnwidth}{!}{
    \renewcommand{\arraystretch}{1.2} 
    \setlength{\tabcolsep}{5pt}      
    \begin{tabular}{lccccccccc}
        \toprule
        \textbf{Key Factors} & \textbf{[9]} & \textbf{[10]} & \textbf{[12]} & \textbf{[15]} & \textbf{[17]} & \textbf{[18]} & \textbf{[19]} & \textbf{[20]} & \textbf{This Paper} \\
        \midrule
        UC-CFmMIMO system & $\times$ & $\times$ & $\checkmark$ & $\times$ & $\checkmark$ & $\times$ & $\times$ & $\times$ & $\checkmark$ \\
        3GPP-3D channel modeling & $\times$ & $\times$ & $\times$ & $\times$ & $\times$ & $\times$ & $\times$ & $\times$ & $\checkmark$ \\
        Frequency resource allocation & $\times$ & $\times$ & $\times$ & $\times$ & $\checkmark$ & $\times$ & $\times$ & $\checkmark$ & $\checkmark$ \\
        Next-generation multiple access (NGMA) & $\times$ & $\times$ & $\times$ & $\times$ & $\times$ & $\checkmark$ & $\times$ & $\checkmark$ & $\checkmark$ \\
        Terrestrial vehicular networks & $\times$ & $\times$ & $\times$ & $\times$ & $\times$ & $\checkmark$ & $\checkmark$ & $\checkmark$ & $\checkmark$ \\
        Dense networks & $\times$ & $\times$ & $\checkmark$ & $\times$ & $\times$ & $\checkmark$ & $\times$ & $\times$ & $\checkmark$ \\
        Multi-objective optimization problem formulation & $\times$ & $\times$ & $\times$ & $\times$ & $\times$ & $\times$ & $\checkmark$ & $\times$ & $\checkmark$ \\
        Bandwidth (\(\geq\)20 MHz) & $\times$ & $\times$ & $\times$ & $\checkmark$ & $\checkmark$ & $\times$ & $\times$ & $\times$ & $\checkmark$ \\
        Multisubband system (\(\geq\) 50 subbands) & $\times$ & $\times$ & $\times$ & $\times$ & $\checkmark$ & $\times$ & $\times$ & $\times$ & $\checkmark$ \\
        \bottomrule
    \end{tabular}
    } 
\end{table}


\section{System Model Description}
\label{sec:systemDescription}
\begin{figure}[!t]
\centering
\begin{center}
    \includegraphics[width=0.45\textwidth, trim=0cm 0cm 0cm 0cm, clip]{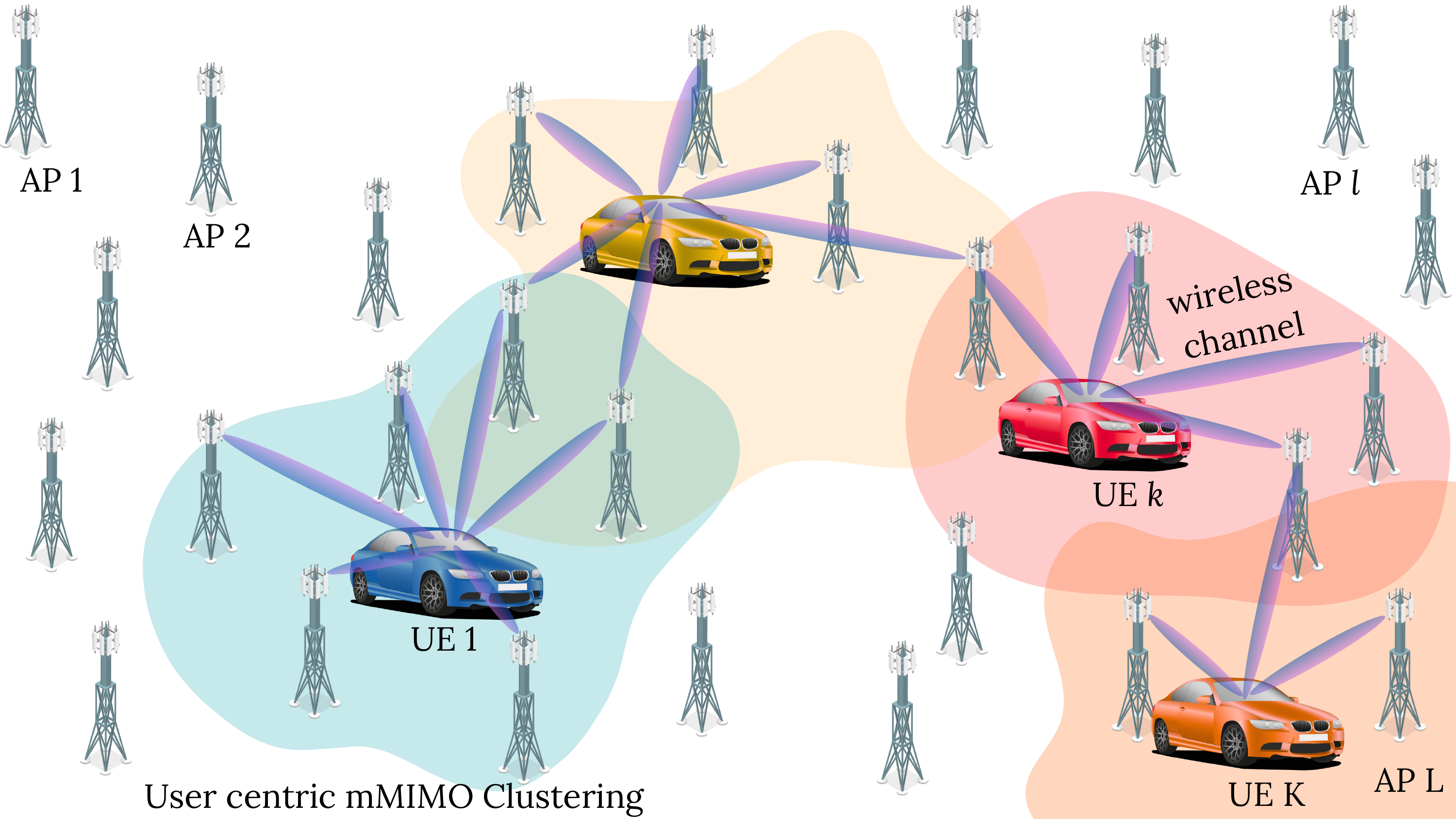}
\end{center}
\caption{Illustration of UC mMIMO clustering in a CF network architecture.}
\label{fig:CarUCmMIMO}
\end{figure}

In our system, we investigate a downlink UC-CFmMIMO setup (figure \ref{fig:CarUCmMIMO}), consisting of \( L \) APs, each equipped with \( N \) antennas. These APs are interconnected to a central processing unit (CPU) via fronthaul links, which coordinate data detection centrally. The APs provide service to \( K \) single-antenna UEs distributed randomly within a predefined geographical area \( R \). To model the communication links between the APs and UE, we employ QuaDRiGa, a widely used framework for simulating realistic wireless propagation conditions. QuaDRiGa extends conventional geometry-based stochastic models by incorporating spatial consistency, temporal evolution, and realistic antenna patterns, making it particularly well suited for multi-antenna and massive MIMO systems. It captures key propagation characteristics such as path loss, shadowing, small-scale fading, and Doppler shifts, ensuring a comprehensive representation of channel dynamics. Additionally, QuaDRiGa incorporates measurement-based parameterization, enabling it to replicate diverse deployment scenarios. This allows for a detailed and reliable analysis of the modeled system under varying propagation conditions.
The system’s total bandwidth, denoted by \(B\), is partitioned into \( S \) subbands, indexed as \( s \in \{1, 2, \ldots, S\} \). We represent the frequency-dependent channel vector between the \( l^{th} \) AP and the \( k^{th} \) UE on subband \( s \) as \( \mathbf{h}_{k,l,s} \). For ease of analysis, we aggregate these channel vectors from all APs to a single UE into the collective channel \( \mathbf{h}_{k,s} \), expressed as \( \mathbf{h}_{k,s} = [\mathbf{h}_{k,1,s}, \ldots, \mathbf{h}_{K,L,s}] \). We assumed that accurate channel state information for all UEs is available at the CPU, facilitated by advanced channel estimation techniques, although the specifics of these techniques are beyond the scope of this paper.

In focusing on frequency resource allocation, we implement a dynamic cooperation clustering framework as introduced in \cite{equalpower1}, enhanced by the AP selection technique outlined in \cite{APselec1}. This method, applied before the communication phase, determines the specific subset of APs responsible for serving each UE. Following this process, the resource allocation process distributes the \( K \) UEs across the \( S \) subbands, producing the assignment matrix \( \mathcal{A} \), where \( \sum_{s=1}^{S} \sum_{k=1}^{K} \mathcal{A}_{sk} = K \). 
The received signal for UE-\( k \) on subband \( s \) is:
\begin{equation} \label{eq2}
y_k = \sum_{l=1}^{L} \mathbf{h}_{k,l,s} \mathbf{x}_{l,s} + n_{k,s},
\end{equation}
where \( n_{k,s} \sim \mathcal{N}(0, \sigma^2) \) represents additive Gaussian noise, and \( \mathbf{x}_{l,s} \) is the transmitted signal from AP \( l \) on subband \( s \).

To define the serving APs, we introduce \( \mathbf{D}_{k,l} \), an \( N \times N \) matrix that is \( \mathbf{I}_N \) if UE-\( k \) is served by AP-\( l \), and \( \mathbf{0}_{N \times N} \) otherwise. The combined \( NL \times NL \) block diagonal matrix for all APs is:
\(
\mathbf{D}_{k} = \mathbf{D}_{k,1} \oplus \mathbf{D}_{k,2} \oplus \cdots \oplus \mathbf{D}_{k,L}
\)\footnote{\small The operation \( \oplus \) constructs an \( NL \times NL \) block diagonal matrix, placing \( \mathbf{D}_{k,l} \) along the diagonal for \( l = 1, 2, \ldots, L \).}.

The transmitted signal \( \mathbf{x}_{l,s} \) is:
\begin{equation} \label{eq3}
\mathbf{x}_{l,s} = \sum_{i \in \mathcal{A}_{s}} \mathbf{D}_{i,l} \mathbf{w}_{i,l} a_i,
\end{equation}
where \( a_i \) is the data signal for UE-\( i \), \( \mathbf{w}_{i,l} \) is the \( N \times 1 \) precoding vector for UE-\( i \), and \( i \in \mathcal{A}_{s} \) refers to the index of the UE assigned to subband \( s \) in the assignment matrix \( \mathcal{A} \). Thus, \( \mathcal{A}_{s} \) is a subset of \( \mathcal{A} \) representing the UEs assigned to subband \( s \).

Substituting into equation \eqref{eq2}, the received signal for UE-\( k \) on subband \( s \) becomes:
\begin{equation}
y_k = \mathbf{h}_{k,s} \sum_{i \in \mathcal{A}_{s}} \mathbf{D}_{i} \mathbf{w}_{i} a_i + n_{k,s}.
\end{equation}

The signal-to-noise-plus-interference ratio (SINR) for UE-\( k \) is:
\begin{equation}
\text{SINR}_k = \frac{|\mathbf{h}_{k,s} \mathbf{D}_{k} \mathbf{w}_{k}|^2}
{\sum_{i \in \mathcal{A}_{s} \backslash k} |\mathbf{h}_{k,s} \mathbf{D}_{i} \mathbf{w}_{i}|^2 + \sigma^2},
\end{equation}
leading to the SE:
\begin{equation} \label{SEEq}
\eta_k = \log_2 (1 + \text{SINR}_k).
\end{equation}

The design of the precoding vector considers the interference predominantly originating from a select group of nearby UEs, defined by the set \( \mathcal{C}_k \):
\begin{equation}
\mathcal{C}_k = \{i : \mathbf{D}_i \mathbf{D}_k \neq \mathbf{0}_{N \times N}\}.
\end{equation}

Taking into account the subband allocation, the zero-forcing (ZF) precoding vector is tailored to minimize interference for UEs sharing the same APs and subbands. The relevant subset of UEs, \( \mathcal{C}_{k,s} \), is incorporated into the channel matrix:
\begin{equation}
\mathbf{H}_{k,s} = [\mathbf{h}_{k,s}; \mathbf{h}_{i,s} : i \in \mathcal{C}_{k,s} \backslash \{k\}],
\end{equation}
where \( \mathcal{C}_{k,s} = \mathcal{A}_{s} \cap \mathcal{C}_k \).

The ZF precoding vector \( \mathbf{w}_k \) is then derived from the equation:
\begin{equation}
\mathbf{H}_{k,s} \mathbf{D}_k \mathbf{w}_k = [1,0,\ldots,0]^T,
\end{equation}
obtained from the first column of:
\begin{equation}
\mathbf{D}_k \mathbf{H}_{k,s}^H (\mathbf{H}_{k,s} \mathbf{D}_k \mathbf{H}_{k,s}^H)^{-1},
\end{equation}
where the superscript $H$ denotes the Hermitian (conjugate transpose) of \( \mathbf{H}_{k,s} \).

The normalized precoding vector \( \mathbf{w}_k \) is scaled to the allocated power \( \rho_k \), computed as:
\begin{equation}
\mathbf{w}_k = \sqrt{\frac{\rho_k}{\| \mathbf{w}_k \|}} \mathbf{w}_k.
\end{equation}
An equal power distribution scheme, as referenced in \cite{equalpower2}, standardizes the power allocation across UEs:
\begin{equation}
\rho_k = \frac{P_{\text{max}}}{\tau_p},
\end{equation}
with $\tau_p$ denoting the pilot sequence length.
\section{Problem Formulation}
\label{sec:problemFormulation}
This paper tackles the challenge of efficiently allocating multiple UEs to a single subband, optimizing limited frequency resources while meeting system constraints. Our approach focuses on maximizing SE, ensuring UE fairness, and minimizing interference in shared subband scenarios. 
We formulate a multi-objective optimization framework using an assignment matrix \( \mathcal{A} \) to dynamically allocate UEs, balancing SE, fairness, and interference reduction.

\subsection{Objective}

\subsubsection{Satisfying a Given User Demand}

Our first objective is to maximize UEs' SE to ensure reliable quality service in a vehicular network. Mathematically, this objective can be represented as follows:
\begin{equation}
\eta_{\text{Total}} = \sum_{s=1}^{S} \sum_{k=1}^{K} \eta(\mathcal{A}_{sk}),
\end{equation}
where \( \eta_{\text{Total}} \) represents the total SE of the system. The term \( \mathcal{A}_{sk} \) refers to the \( k^{\text{th}} \) UE assigned to subband \( s \), and \( \eta(\mathcal{A}_{sk}) \) denotes the SE achieved by that UE on subband \( s \).

\subsubsection{Ensuring an Equitable Distribution Among Users}

To ensure fairness in resource distribution and prevent any UE from being disproportionately advantaged or disadvantaged by the frequency-specific characteristics of the channel, we employ the Gini index (\( I_{\text{Gini}}) \) to monitor and manage fairness across the system, as described in \cite{giniI1}:

\begin{equation}
I_{\text{Gini}} = \frac{1}{2K^2 \bar{\eta}} \sum_{s=1}^{S} \sum_{i=1}^{K} \sum_{j=1,j\neq i}^{K} |\eta(\mathcal{A}_{si}) - \eta(\mathcal{A}_{sj})|,
\end{equation}
where \( \bar{\eta} \) is the average SE, and \( \eta(\mathcal{A}_{si}) \) and \( \eta(\mathcal{A}_{sj}) \) are the SEs of UEs \( i \) and \( j \) on subband \( s \), respectively.
We favor the Gini index in our analysis as it captures fairness across the entire UE base, providing a balanced perspective that accounts for disparities among all UEs, rather than focusing solely on the least-advantaged (as in max-min fairness) or proportional allocation.

\subsubsection{Reducing Interference}

Eigenvalue Decomposition (EVD) is used to reduce interference by improving the conditioning of the channel matrix. By maximizing the smallest eigenvalue of the decomposed matrix, we ensure less correlation between UEs that will be allocated in the same subband, which minimizes interference and enhances overall system performance.

Let $\mathbf{H}_s = [\mathbf{h}_{1,s}, \ldots, \mathbf{h}_{K,s}]$ represent the channel matrix for subband $s$. The matrix $\mathbf{D}$ is a block diagonal matrix composed of $\mathbf{D}_k$ for all $K$ UEs, where:

\[
\mathbf{D} = \text{diag}(\mathbf{D}_1, \mathbf{D}_2, \ldots, \mathbf{D}_K).
\]
The eigenvalue decomposition is performed on the matrix \( \mathbf{H}_s^T \mathbf{D} \mathbf{H}_s \) as follows:
\begin{equation}
\mathbf{H}_s^T \mathbf{D} \mathbf{H}_s = \mathbf{V}_s \mathbf{\Lambda}_s \mathbf{V}_s^T,
\end{equation}
where the superscript \( T \) denotes the transpose operation applied to the corresponding matrices \( \mathbf{H}_s \) and \( \mathbf{V}_s \), \( \mathbf{\Lambda}_s \) contains the eigenvalues of \( \mathbf{H}_s^T \mathbf{D} \mathbf{H}_s \), and \( \mathbf{V}_s \) is the matrix of eigenvectors. Through our objective function, we aim to maximize the smallest eigenvalue, ensuring that UEs experiencing the highest interference are assigned subbands with better channel conditions, thereby minimizing interference across the network:
\begin{equation}
\max \lambda_{\min} \quad \text{where} \quad \lambda_{\min} = \min_{s \in [1, S]} \left( \min(\mathbf{\Lambda}_s) \right).
\end{equation}

\subsection{Optimization Problem}
The primary objective of this work is to jointly optimize the system's performance by maximizing the total SE, minimizing interference, and promoting fairness in UE resource distribution. This multi-objective optimization problem is formulated as follows:

\begin{equation}
    \max_{\mathcal{A}} \left( w_{\eta} \times \eta_{Total} + w_{\text{EVD}} \times \lambda_{\min} - w_{\text{Gini}} \times I_{\text{Gini}} \right),
    \label{eq:objective}
\end{equation}

Subject to:

\begin{subequations}
\begin{align}
    \hspace{0.5cm} \textbf{C1:} \quad & \sum_{k=1}^{K} \rho_k \leq \rho_{\text{max}}, 
    \label{eq:C1} \\
    \hspace{0.5cm} \textbf{C2:} \quad & \begin{aligned}
    \eta(\mathcal{A}_{sk}) \geq \eta_{\text{th}}, \quad \forall s \in \{1, \dots, S\}, \\
    \forall k \in \{1, \dots, K\},
    \end{aligned}
    \label{eq:C2} \\
    \hspace{0.5cm} \textbf{C3:} \quad & \sum_{s=1}^{S} |\mathcal{A}_{sk}| \leq 1, \quad \forall k \in \{1, \dots, K\},
    \label{eq:C3} \\
    \hspace{0.5cm} \textbf{C4:} \quad & \sum_{s=1}^{S} |\mathcal{A}_{s}| \leq K. 
    \label{eq:C4}
\end{align}
\end{subequations}

The optimization problem in \eqref{eq:objective} is designed to achieve a balance between three important objectives: \textbf{\textit{i}}) maximizing SE, \textbf{\textit{ii}}) minimizing interference, and \textbf{\textit{iii}}) ensuring fairness. Weights \( w_{\eta} \), \( w_{\text{EVD}} \), and \( w_{\text{Gini}} \) are utilized to reflect the relative importance of each objective. To maintain system energy efficiency, constraint \eqref{eq:C1} limits the total power allocated to all UEs, capping it at \( \rho_{\text{max}} \). Additionally, constraint \eqref{eq:C2} guarantees that each UE achieves a minimum SE, denoted by \( \eta_{\text{th}} \) across all subbands, ensuring a consistent level of service quality throughout the network. Further, constraint \eqref{eq:C3} prevents any UE from being assigned to more than one subband, thereby ensuring a fair and efficient distribution of resources. Lastly, constraint \eqref{eq:C4} ensures that UEs are assigned to subbands based on the available frequency resources and the number of UEs seeking to connect, without exceeding the total number of available UEs. This formulation seamlessly integrates these objectives and constraints, optimizing system performance while adhering to practical considerations such as power limits, UE allocation, and service quality.


\section{Adaptive Resource Allocation for the System}
\label{sec:adaptiveSolutions}

To efficiently solve the complex optimization problem of frequency resource allocation in UC-CFmMIMO systems, adaptive strategies are essential to balance SE, fairness, and interference mitigation. Given the dynamic nature of vehicular networks, traditional allocation methods often fall short due to their limited adaptability and scalability. This section explores advanced methods for resource allocation, including metaheuristic optimization, model-free reinforcement learning, and hybrid approaches. These methods are designed to address the multifaceted challenges posed by dynamic environments while achieving high-performance gains. 

\subsection{Metaheuristic Solution}
Metaheuristic algorithms are widely recognized for their effectiveness in tackling complex optimization problems in wireless networks by efficiently exploring vast and intricate solution spaces \cite{meta}. These techniques, inspired by a diverse range of natural, biological, and physical phenomena, provide robust global search capabilities through a trade-off between exploration and exploitation. Despite their versatility, selecting an appropriate metaheuristic for a given problem requires careful consideration of the problem's structure and constraints. In the case of CFmMIMO systems within vehicular networks, frequency resource allocation presents a particularly challenging optimization landscape. The problem exhibits a high degree of irregularity, with abrupt variations in the search space caused by channel frequency dependency, fluctuating channel conditions, and evolving interference patterns. These factors contribute to a non-convex, non-differentiable, and highly discontinuous objective function, where conventional gradient-based methods often fail. Their reliance on smooth gradients makes them prone to convergence issues or stagnation in local optima, rendering them ineffective for such a complex problem. Given these challenges, the AO, introduced in 2021 \cite{AO}, has emerged as a compelling solution due to its bio-inspired hunting strategies that effectively balance global exploration and local exploitation. AO operates through four primary mechanisms: exploratory movements to scan the search space, targeted exploitation of promising regions, diagonal movements to enhance search coverage, and random perturbations to escape local optima. These adaptive mechanisms make AO particularly well-suited for optimization landscapes characterized by frequency-selective fading, rapid channel fluctuations, and unpredictable interference variations, as encountered in our study. Unlike traditional metaheuristics, AO’s structured multi-phase search process allows it to efficiently locate optimal solutions while maintaining robustness against the stochastic and frequency-dependent nature of CFmMIMO-based resource allocation in vehicular networks.

\subsection{ Reinforcement Learning-Based Solution}

Machine learning has gained significant attention in wireless communication for its ability to handle complex decision-making problems where traditional model-based approaches fall short. Learning-based methods can be broadly categorized into supervised, unsupervised, and reinforcement learning (RL) paradigms. Supervised learning relies on labeled datasets to train models, making it effective for classification and regression tasks but impractical for problems where labeled data is unavailable or costly to obtain. Unsupervised learning, on the other hand, identifies patterns in unlabeled data but lacks a structured mechanism for decision-making in dynamic environments. Reinforcement Learning (RL) offers a distinct advantage by enabling an agent to learn optimal strategies through direct interaction with the environment, eliminating the need for predefined datasets \cite{rlSurvey}. In RL, an agent sequentially makes decisions, receives feedback in the form of rewards, and continuously refines its policy to maximize long-term performance. This trial-and-error approach is particularly effective for problems where system behavior is influenced by multiple, interdependent factors. 

As discussed before, UC-CFmMIMO-based frequency resource allocation presents a complex optimization challenge, shaped by non-stationary network conditions. Traditional model-driven optimization techniques struggle to accurately capture these fluctuations, making it difficult to develop static allocation policies. RL provides a model-free learning-based framework that iteratively improves allocation decisions by leveraging past experiences and observed environmental conditions. This capability makes RL particularly well-suited for frequency allocation strategies, where channel variations and interference significantly impact performance. To provide a structured understanding of RL’s role in this study, the following subsections introduce the fundamentals of RL and outline its specific application to CFmMIMO resource allocation.

\subsubsection{Basics of Model-Free RL}
Model-free RL is a category of reinforcement learning in which an agent learns decision policies solely based on observed interactions, without requiring an explicit mathematical model of system dynamics. The agent operates through a sequence of states, actions, and rewards: it observes the current state, selects an action, and receives feedback in the form of a reward, which guides its learning process. Over time, the agent improves its policy by maximizing cumulative rewards while balancing exploration (trying new actions) and exploitation (leveraging known actions).  

Model-free RL has been successfully applied in wireless networks, particularly for resource allocation tasks where system dynamics are too complex to be explicitly modeled \cite{rlBasics}. Techniques such as Q-Learning and Deep Q-Networks (DQN) have demonstrated strong potential in optimizing spectrum allocation by enabling adaptive decision-making under uncertain conditions.

\subsubsection{Application of RL to UC-CFmMIMO Frequency Allocation}
In this study, RL is employed to optimize multi-UE frequency resource allocation in CFmMIMO systems. Unlike static optimization methods, RL provides a framework for iteratively refining frequency allocation strategies based on observed system conditions. The goal is to develop allocation policies that maximize SE while mitigating interference and ensuring fairness among UEs. Specifically, model-free RL is leveraged to adjust subband allocations by analyzing past allocations and assessing their effectiveness in minimizing interference and optimizing SE. Through this learning-based approach, RL-based allocation strategies can enhance fairness, SE, and interference mitigation in CFmMIMO deployments. By learning from accumulated interactions, RL enables a structured decision-making process that accounts for the stochastic and frequency-dependent nature of CFmMIMO resource allocation.

\subsection{Hybrid Solution}

Effectively addressing the challenges of frequency resource allocation in UC-CFmMIMO systems requires leveraging multiple optimization paradigms. While metaheuristic algorithms provide strong global search capabilities, they lack adaptive decision-making based on system feedback. Conversely, RL excels at policy optimization through experience-driven learning but can struggle with convergence in highly complex search spaces. To capitalize on the strengths of both approaches, this study proposes a hybrid framework that integrates the AO with DDPG-based AC-RL, as outlined in Algorithm \ref{alg:hybrid_AO_AC}. The proposed hybrid approach utilizes AO’s exploratory mechanisms to enhance the action space exploration of AC-RL, thereby improving sample efficiency and preventing premature convergence to suboptimal solutions. Specifically, AO is employed to generate exploratory actions that guide the RL agent in selecting more efficient frequency allocations, ensuring that the search space is effectively covered. Meanwhile, AC-RL refines these decisions over multiple interactions by learning optimal policies through reward-based adaptation. As detailed in Algorithm \ref{alg:hybrid_AO_AC}, the hybrid framework operates by first initializing the Actor and Critic networks alongside their respective target networks. During each episode, AO assists in generating exploratory actions, which are then executed within the RL environment. The observed state transitions and rewards are stored in a replay buffer, from which mini-batches are sampled to update the Critic network using value function approximation. The Actor network is subsequently updated based on the policy gradient, while target networks undergo soft updates to stabilize training. By combining AO’s structured global search with AC-RL’s adaptive learning capabilities, the hybrid approach effectively balances exploration and exploitation, leading to improved convergence of the objective function given in \eqref{eq:objective}. This, in turn, enhances SE, mitigates interference, and ensures fairness in UC-CFmMIMO-based frequency allocation. This integration not only enhances solution quality but also reduces the training time required for RL-based optimization by guiding the agent toward promising regions of the search space. The following sections present a detailed performance evaluation of the proposed hybrid framework, comparing its effectiveness against standalone AO and AC-RL approaches.

\begin{algorithm}[]
\caption{Hybrid AO with AC-RL}
\label{alg:hybrid_AO_AC}
\begin{algorithmic}[1]
\STATE \textbf{Initialize:} Actor network $\mu(s|\theta^\mu)$, Critic network $Q(s,a|\theta^Q)$, 
\STATE Initialize Target networks: $\theta^{\mu'} \leftarrow \theta^{\mu}$, $\theta^{Q'} \leftarrow \theta^{Q}$
\STATE Initialize replay buffer $\mathcal{D}$

\FOR{each episode}
    \STATE Reset environment and obtain initial state $s_0$
    
    \FOR{each timestep $t$}
        \STATE \textbf{Exploration:} Use \textbf{AO} to generate exploratory action $a_t$
        \STATE Execute action $a_t$ in the environment, observe reward $r_t$ and next state $s_{t+1}$
        \STATE Store transition $(s_t, a_t, r_t, s_{t+1})$ in replay buffer $\mathcal{D}$
        
        \STATE \textbf{Critic Update:}
        \STATE Sample a mini-batch $(s, a, r, s')$ from $\mathcal{D}$
        \STATE Compute target value: $y_t = r + \gamma Q'(s', \mu'(s')|\theta^{Q'})$
        \STATE Compute loss: $L = (y_t - Q(s,a|\theta^Q))^2$
        \STATE Update Critic network $\theta^Q$ by minimizing $L$
        
        \STATE \textbf{Actor Update:}
        \STATE Compute policy gradient $\nabla_{\theta^\mu} J = \mathbb{E}[\nabla_a Q(s,a|\theta^Q) \nabla_{\theta^\mu} \mu(s|\theta^\mu)]$
        \STATE Update Actor network $\theta^\mu$ using gradient ascent
        
        \STATE \textbf{Target Networks Soft Update:}
        \STATE $\theta^{Q'} \leftarrow \tau \theta^Q + (1 - \tau) \theta^{Q'}$
        \STATE $\theta^{\mu'} \leftarrow \tau \theta^\mu + (1 - \tau) \theta^{\mu'}$
    \ENDFOR
\ENDFOR
\end{algorithmic}
\end{algorithm}
\subsection{Complexity Analysis}
The computational complexity of the proposed solutions AO, AC-RL, and HRLM (Hybrid RL with AO integration) reflects their distinct approaches to solving the resource allocation problem, each offering a balance between precision, adaptability, and computational feasibility. AO (Alternating Optimization) is a structured, deterministic method where the complexity is driven by iterative function evaluations and matrix operations, making it highly effective for structured optimization problems but computationally demanding as it scales quadratically with the population size \( N \). Its overall complexity is \( O(T \times N^2 \times K \times S \times L) \), where \( T \) is the number of iterations, \( K \) is the number of UEs, \( S \) is the number of subbands, and \( L \) is the number of APs. On the other hand, AC-RL introduces an adaptive learning-based approach where training complexity is dominated by forward and backward passes through deep neural networks. This allows the model to dynamically improve decision-making but requires substantial computational resources, with a complexity of \( O(T \times B \times D \times N_{\text{neurons}}^2) \), where \( B \) is the batch size, \( D \) is the number of layers, and \( N_{\text{neurons}} \) represents the number of neurons per layer. Recognizing the strengths of both methods, HRLM leverages the AO to enhance exploration within reinforcement learning, striking a balance between structured optimization and adaptive decision-making. However, this added sophistication introduces an additional complexity term \( O(P \times I \times K \times S) \), where \( P \) represents the population size in AO, and \( I \) is the number of iterations per optimization step, leading to an overall complexity of \( O(T \times B \times D \times N_{\text{neurons}}^2) + O(P \times I \times K \times S) \). While AO provides a mathematically rigorous solution with high computational demands, AC-RL enables real-time adaptability, and HRLM strategically merges the best of both worlds, maximizing efficiency while keeping computational overhead in check. This hybrid approach stands out as a powerful and scalable solution, capable of navigating the trade-off between performance and computational cost, making it highly competitive in tackling complex optimization challenges.

\section{Simulation Setup and Data Generation}
\label{sec:simulationSetup}

\begin{figure*}[!t]
\centering
\begin{center}
    \includegraphics[width=0.9\textwidth, trim=0cm 0cm 0cm 0cm, clip]{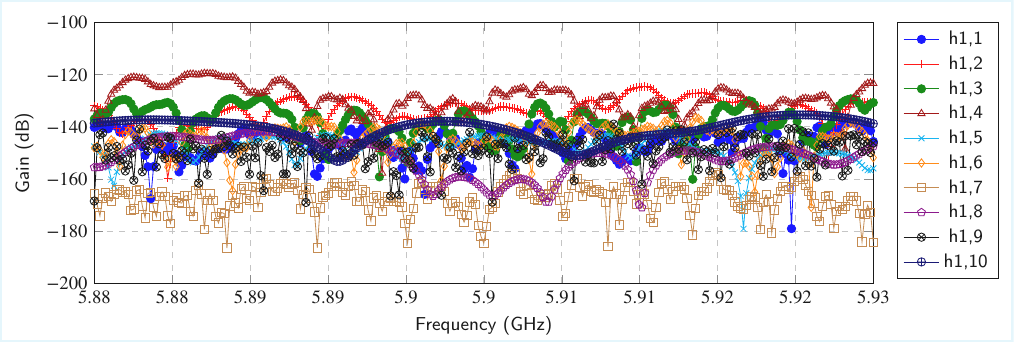}
\end{center}
\caption{Channel frequency responses of the 10 AP-UE pair links.}
\label{fig:cfr_output}
\end{figure*}


To emulate a real-life communication scenario, we designed a simulation framework to generate channel frequency responses (CFRs) for the deployed AP and UE pairs. These CFRs form the foundation for the performance evaluations presented in the subsequent sections, enabling a realistic assessment of system performance in a dense urban environment.
The simulation is conducted over a \( 1 \, \mathrm{km} \times 1 \, \mathrm{km} \) area, with \( L = 100 \) APs uniformly distributed and positioned at a height of \( h_{\text{AP}} = 12.5 \, \mathrm{m} \). The \( K = 40 \) UEs, modeled as vehicles, are randomly placed within the area at \( h_{\text{UE}} = 1.5 \, \mathrm{m} \). The system operates at a carrier frequency of \( f_c = 5.9 \, \mathrm{GHz} \), which corresponds to the allocated frequency for vehicular communications in 5G and B5G networks \cite{vehiculefreq1, vehiculefreq2, vehiculefreq3}, with a total bandwidth of \( B = 50 \, \mathrm{MHz} \). This bandwidth is divided into 277 subbands, each consisting of 12 subcarriers with a spacing of \( B_\text{RB} = 180 \, \mathrm{kHz} \), ensuring compliance with modern vehicular communication standards.

The deployment parameters for the simulation are summarized in Table~\ref{tab:sim_params}, encompassing the physical dimensions, system bandwidth, antenna configurations, and UE placement characteristics. This detailed configuration ensures that the generated CFRs accurately capture the frequency-selective and time-varying nature of wireless channels in urban environments. By representing the frequency-domain channel coefficients for each AP-UE pair across all subcarriers and time instances, CFRs provide a comprehensive view of propagation effects. Figure \ref{fig:cfr_output} illustrates an example, depicting the channel gain (in dB) across the simulated frequency range. The figure provides a representative view of channel responses for 10 AP-UE links, specifically focusing on the first 10 UEs connected to AP 1 (from \( h_{1,1} \) to \( h_{1,10} \)) to enhance clarity and readability. The observed variations in channel gain over the 5.88–5.92 GHz range highlight the highly frequency-selective nature of the modeled channels, influenced by multipath fading, interference, and frequency-dependent attenuation. These fluctuations are characteristic of dense vehicular communication networks, where rapid changes in propagation conditions significantly impact performance.

\begin{table}[h!]
    \centering
    \caption{QuaDRiGa Simulation Parameters.}
    \label{tab:sim_params}
    \resizebox{\columnwidth}{!}{
    \begin{tabular}{llll}
        \hline
        \textbf{Parameter} & \textbf{Description} & \textbf{Value} & \textbf{Reference} \\ \hline
        \( f_c \) & Central Frequency & \( 5.9 \, \mathrm{GHz} \) & \cite{vehiculefreq3} \\ 
        \( B \) & Total bandwidth & \( 50 \, \mathrm{MHz} \) & \cite{5GStandard1} \\ 
        \( B_\text{RB} \) & Frequency resource block & \( 180 \, \mathrm{kHz} \) & \cite{5GStandard1} \\ 
        UCA & Uniform Circle Array & AP antennas type & \cite{UCA} \\ 
        ULA & Uniform Linear Array & UE antennas type & \cite{ULA} \\ 
        \( R \) & Area of coverage & \( 2 \, \mathrm{km} \times 2 \, \mathrm{km} \) & \cite{CF1} \\ 
        \( L \) & Number of APs & \( 100 \) & \cite{setupAPpar} \\ 
        \( N \) & Number of antennas per AP & \( 4 \) & \cite{setupAPpar} \\ 
        \( K \) & Number of UEs & \( 40 \) & \cite{setupAPpar} \\ 
        \( N_{\text{UE}} \) & Number of antennas per UE & \( 1 \) & \cite{CF1} \\ 
        \( h_{\text{AP}} \) & AP height & \( 12.5 \, \mathrm{m} \) & \cite{setupPar1L} \\ 
        \( h_{\text{UE}} \) & UE height & \( 1.5 \, \mathrm{m} \) & \cite{setupPar1L} \\ 
        \( \tau_p \) & Pilot sequence & \( 10 \) & \cite{setupPar1L} \\ 
        \( P_{\max} \) & DL transmit power per AP & \( 200 \, \mathrm{mW} \) & \cite{setupPar1L} \\ 
        \( S \) & Total number of simulated subbands & \( 277 \) & - \\ 
        3GPP\_38.901\_UMi  & Deployment environment & Urban & - \\ \hline
    \end{tabular}
    } 
\end{table}


\section{Implementation of Reinforcement Learning-Based Solutions}
\label{sec:RL_implementation}
This section details the implementation of the RL frameworks designed for the previously introduced problem. First, we present the AC-DDPG model, which serves as the initial RL-based solution for optimizing frequency resource allocation in CFmMIMO systems. Next, we introduce its hybridization with the AO, leveraging AO's global search capability to enhance the exploration process of the AC architecture. Finally, we outline the hyperparameter optimization strategy, which employs a random search approach to fine-tune the learning parameters and improve performance.

\begin{figure}[!t]
\centering
\begin{center}
    \includegraphics[width=0.5\textwidth, trim=2.5cm 2.5cm 2cm 2.3cm, clip]{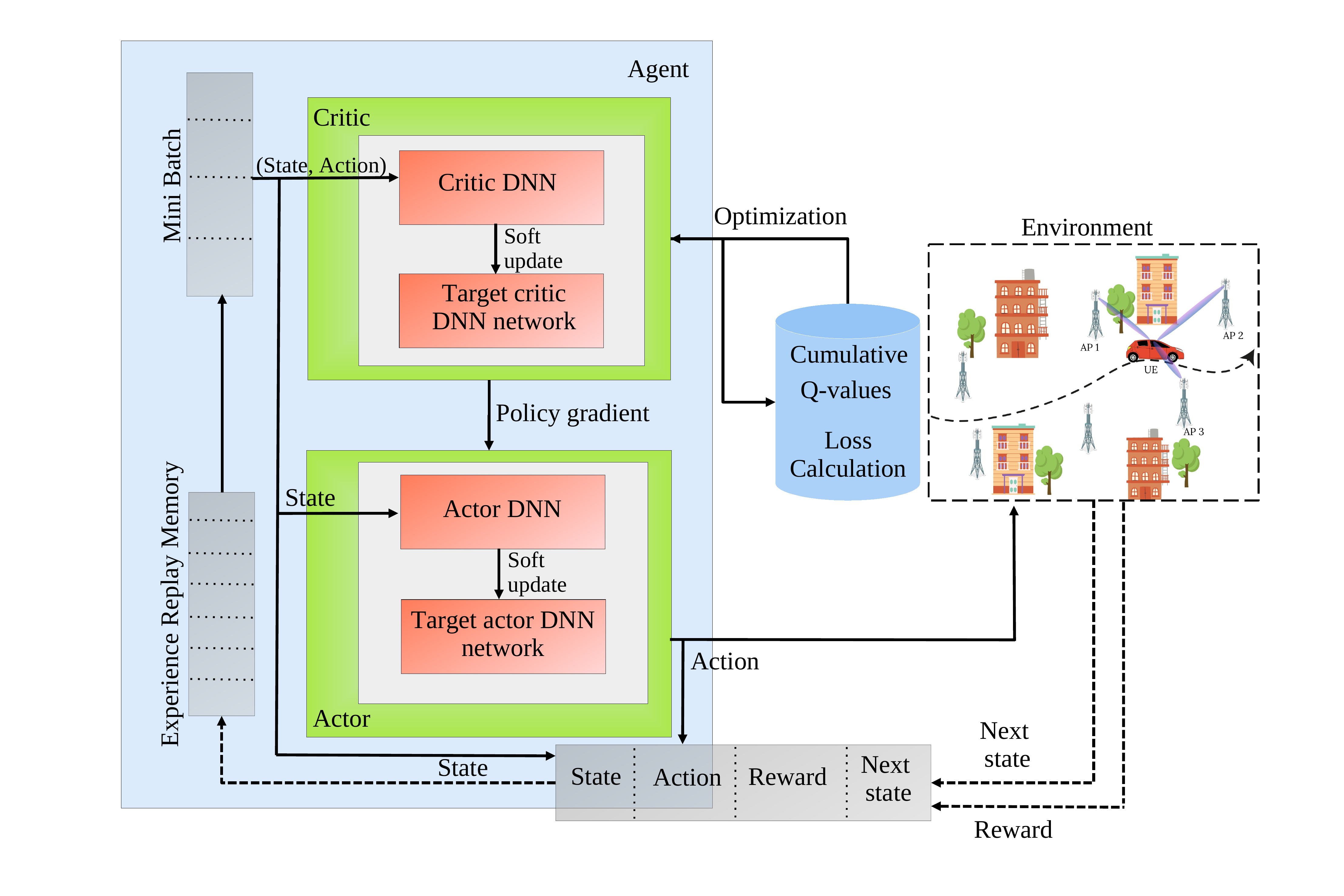}
\end{center}
\caption{DDPG-based RLM framework.}
\label{fig:DDPG_RLM}
\end{figure}

\subsection{DDPG Learning Approach}
The DDPG algorithm is an off-policy reinforcement learning approach specifically designed for high-dimensional continuous action spaces \cite{DDPG}. By integrating off-policy learning with a deterministic approach, the DDPG algorithm ensures both stability and efficiency in decision-making, where the deterministic aspect enables precise action selection while off-policy learning enhances sample efficiency and adaptability in dynamic environments. To establish an effective learning process, the proposed framework follows a structured interaction cycle in which the RL agent perceives its current state \(s_t\), selects an action \(a_t\), receives a reward \(r_t\) reflecting the effectiveness of its action, and transitions to a new state \(s_{t+1}\). This iterative trial-and-error approach allows the agent to refine its policy and optimize decision-making while maximizing cumulative rewards. Since DDPG employs an off-policy strategy, it further enhances learning efficiency by allowing the agent to leverage past experiences that were not necessarily collected under the current policy. This characteristic broadens the learning scope, improves convergence, and refines the model’s overall decision-making capabilities, reinforcing the advantages of off-policy learning within the framework.

To effectively implement this off-policy learning mechanism, the training process incorporates a replay buffer that stores past experiences \((s_t, a_t, r_t, s_{t+1})\), enabling the agent to learn from a diverse set of samples rather than being limited to recent transitions. This mechanism improves sample efficiency and mitigates issues related to correlated updates. Moreover, the use of target networks stabilizes learning by decoupling policy updates from the rapidly fluctuating main networks.

Building on the previously discussed deterministic policy aspect, DDPG further reinforces decision-making precision by leveraging this mechanism alongside its off-policy strategy. Unlike stochastic approaches, where actions are sampled from a probability distribution, the DDPG actor network deterministically maps states to specific actions \(a_t = \pi_\theta(s_t)\), where \( \pi_\theta \) represents the policy function parameterized by \( \theta \), mapping a given state \( s_t \) to its corresponding action \( a_t \). This deterministic mapping ensures consistency in action selection, which is particularly beneficial for optimization problems requiring precise control over resource allocation. Additionally, this approach is computationally efficient, reducing unnecessary exploration noise and facilitating faster convergence to an optimal policy. Figure \ref{fig:DDPG_RLM} illustrates the operational flow of the DDPG learning mechanism within the proposed framework.
\begin{figure}[!t]
\centering
\begin{center}
    \includegraphics[width=0.48\textwidth, trim=0cm 0cm 0cm 0cm, clip]{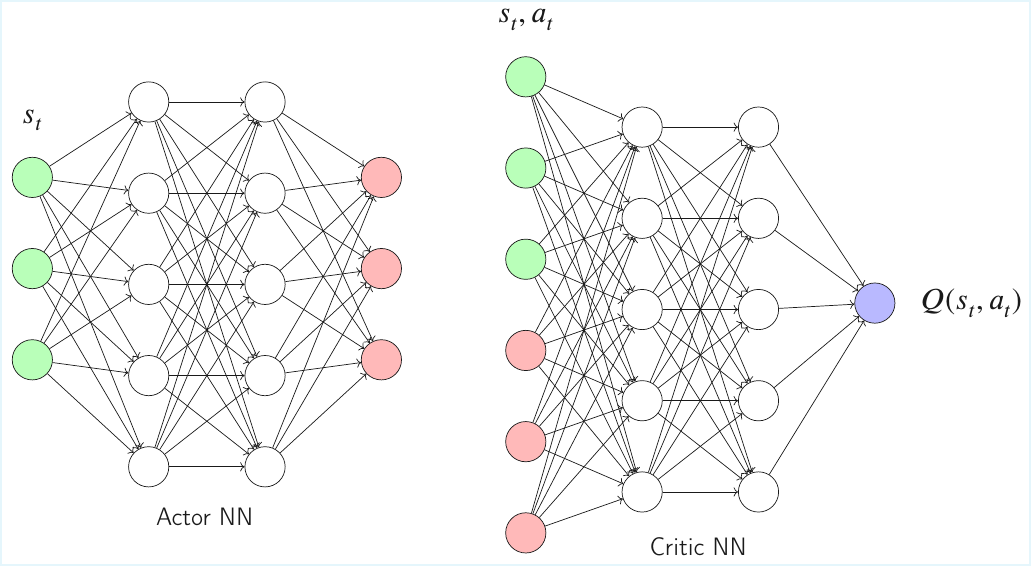}
\end{center}
\caption{Architecture of Actor and Critic deep neural networks (DNNs).}

\label{fig:actor_critic_nn}
\end{figure}
\subsubsection{Neural Network Architecture for Actor-Critic Framework}
The reinforcement learning agent operates within an environment that models the dynamic resource allocation scenario. As previously introduced, the learning process is driven by the interaction between two fundamental components: \textbf{\textit{i)}} \textit{the actor network}, responsible for policy learning, and \textbf{\textit{ii)}} \textit{the critic network}, which evaluates the quality of selected actions by estimating the state-action value function \(Q(s_t, a_t)\). This Q-value representation provides an expected cumulative reward for each state-action pair, enabling the optimization of the policy based on value-driven learning.

The actor and critic networks are implemented using fully connected deep neural networks, as shown in figure \ref{fig:actor_critic_nn}. The actor network receives the state \(s_t\) as input, representing system conditions, including AP-UE clustering, channel characteristics, and the allocation matrix. It consists of two fully connected hidden layers with ReLU activations, allowing the extraction of high-dimensional features. The output layer maps the processed state to a deterministic action \(a_t\) through a softmax layer, ensuring normalized allocation decisions.

The critic network, on the other hand, takes as input the concatenated state-action pair \((s_t, a_t)\). Similar to the actor network, it contains two fully connected hidden layers with ReLU activations, enabling a non-linear approximation of the Q-value function. The output layer produces the Q-value \(Q(s_t, a_t)\), representing the expected cumulative reward for a given input. This structured neural network design ensures stable and efficient policy evaluation within the reinforcement learning framework.

\begin{figure*}[!t]
\centering
\begin{center}
    \includegraphics[width=0.8\textwidth, trim=0cm 0cm 0cm 13cm, clip]{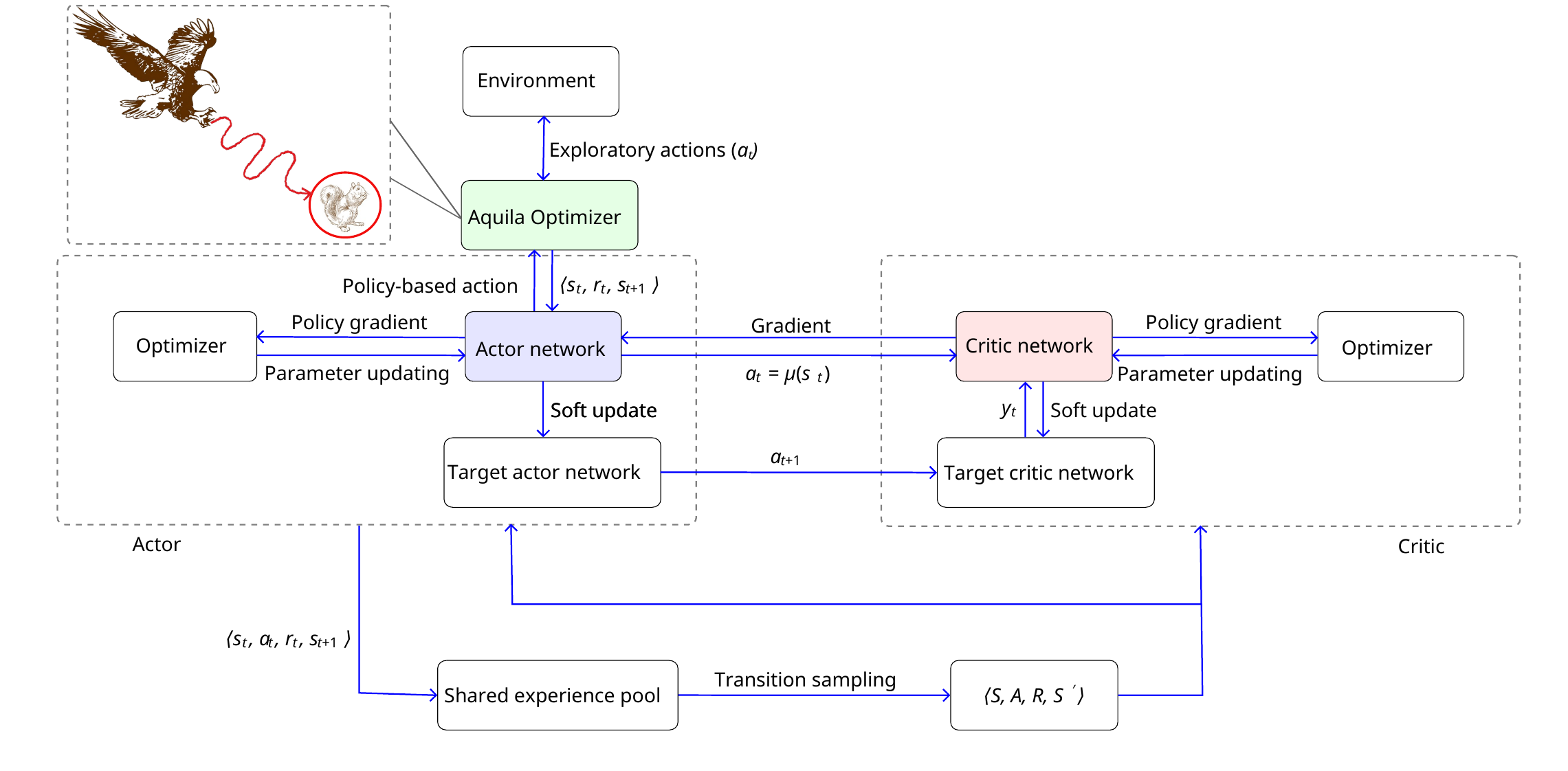}
\end{center}
\caption{Proposed HRLM framework.}

\label{fig:hybrid_ddpg_ao}
\end{figure*}

\subsubsection{Training Workflow and Equations}
The training process follows the Actor-Critic framework and includes the following steps:
\begin{enumerate}[label=\roman*.]
    \item \textbf{Critic Network Update:}
    The critic minimizes the temporal difference (TD) error to improve Q-value estimates:
    \[
    \mathcal{L}_{critic} = \mathbb{E}\left[\left(y_t - Q(s_t, a_t)\right)^2\right],
    \]
    where the target Q-value \(y_t\) is computed as:
    \[
    y_t = r_t + \gamma Q'(s_{t+1}, a'_{t+1}),
    \]
    where \( Q'(s_{t+1}, a'_{t+1}) \) is the target Q-value, computed using the target critic network \( Q' \), and \( a'_{t+1} \) is the next action, obtained from the target actor network.

    The critic network parameters \(\theta_Q\) are updated using gradient descent with learning rate \(\alpha_Q\):
    \[
    \theta_Q \leftarrow \theta_Q - \alpha_Q \nabla_{\theta_Q} \mathcal{L}_{critic}.
    \]

    \item \textbf{Actor Network Update:}
    The actor is trained to maximize the Q-value of the actions it selects. This is achieved using the policy gradient:
    \[
    \nabla_\theta J = \mathbb{E}\left[\nabla_a Q(s_t, a)\big|_{a = \pi_\theta(s_t)} \nabla_\theta \pi_\theta(s_t)\right].
    \]

    The actor network parameters \(\theta_{\pi}\) are updated using gradient ascent with learning rate \(\alpha_{\pi}\):
    \[
    \theta_{\pi} \leftarrow \theta_{\pi} + \alpha_{\pi} \nabla_{\theta_{\pi}} J.
    \]

    \item \textbf{Target Network Update:}
    Target networks are updated using a soft update mechanism:
    \[
    \theta' \leftarrow \tau \theta + (1 - \tau) \theta',
    \]
    where \(\tau\) is the soft update coefficient (\(0 < \tau < 1\)), \(\theta\) represents the parameters of the current network (actor or critic), and \(\theta'\) denotes the parameters of the corresponding target network.

    \item \textbf{Experience Replay:}
    Transitions \((s_t, a_t, r_t, s_{t+1})\) are stored in a shared replay buffer. Mini-batches are sampled randomly during training, breaking temporal correlations and improving sample efficiency.

    \item \textbf{Exploration and Exploitation:}
    Exploration is facilitated using an epsilon-greedy strategy, where random actions are selected with probability \(\epsilon\). The value of \(\epsilon\) decays over time to favor exploitation of the learned policy.
\end{enumerate}


\subsection{Hybrid DDPG-AO Learning Approach}
Figure \ref{fig:hybrid_ddpg_ao} illustrates the proposed hybrid framework that integrates the AO into the AC-DDPG architecture. In this design, the environment provides the current state \( s_t \), which is processed by both the conventional DDPG components (actor and critic) and the AO module. Specifically, the AO module generates exploratory actions for the actor network, helping it escape local optima and facilitating broader exploration of the action space. These exploratory actions, combined with the actor’s deterministic outputs, are then applied to the environment, yielding a next state \( s_{t+1} \) and a reward \( r_t \).

The agent’s experiences \(\bigl(s_t, a_t, r_t, s_{t+1}\bigr)\) are stored in a shared replay buffer and sampled for training the AC networks. The critic network evaluates the quality of actions via its Q-value function and provides a learning signal for adjusting both actor and critic parameters. To stabilize training, target networks for the actor and critic are updated using a soft update mechanism. At the same time, the AO module continuously refines the exploration path by analyzing the reward signals and state transitions, thereby directing the agent toward more promising actions. This synergy effectively merges AO’s global search capability with DDPG’s value-driven policy refinement, leading to faster convergence and improved performance in complex or high-dimensional tasks.

\subsection{Random Search for DDPG Hyperparameter Tuning}
As part of the model-building process for the DDPG framework, hyperparameter optimization was conducted using a random search approach to efficiently identify configurations that balance learning efficiency and performance. Key hyperparameters were sampled from predefined ranges (see Table~\ref{tab:hyperparam_ranges}), including the actor and critic learning rates (\(\theta_{Q}\), \(\theta_{\pi}\)), discount factor (\(\gamma\)), replay buffer size, mini-batch size, soft update coefficient (\(\tau\)) for the target networks, and the magnitude of exploration noise. These specific intervals were chosen based on typical values used in reinforcement learning literature and practical constraints of the application. A total of 6 random configurations were evaluated, with each configuration trained for a fixed number of episodes and assessed on cumulative rewards, convergence speed, and stability metrics. The best-performing configuration was then selected for final training and evaluation. By systematically exploring diverse hyperparameter sets without exhaustive computation, random search provided a computationally efficient means to uncover near-optimal settings for the AC model, ultimately offering a robust foundation for the subsequent evaluation stages.

\begin{table}[h!]
    \centering
    \caption{Hyperparameter Ranges Used in Random Search.}
    \label{tab:hyperparam_ranges}
    \begin{tabular}{lcc}
        \hline
        \textbf{Hyperparameter} & \textbf{Range} & \textbf{Reference} \\ \hline
        Actor learning rate (\(\theta_{\pi}\)) & \(10^{-5}\) to \(10^{-3}\) & \cite{learningrate} \\ 
        Critic learning rate (\(\theta_{Q}\)) & \(10^{-5}\) to \(10^{-3}\) & \cite{learningrate} \\ 
        Discount factor (\(\gamma\)) & 0.9 to 0.99 & \cite{DiscountFactor} \\ 
        Replay buffer size & \(10^4\) to \(10^6\) & \cite{replaybuffer} \\ 
        Batch size & 32 to 256 & \cite{batchsizeSoftUpdate} \\ 
        Soft update coefficient (\(\tau\)) & 0.001 to 0.01 & \cite{batchsizeSoftUpdate} \\ 
        Exploration noise magnitude & 0.1 to 0.5 & \cite{ExplorationNoise} \\ \hline
    \end{tabular}
\end{table}

\subsection{Performance Analysis and Best Model Selection}
\begin{figure}[!t]
\centering
\begin{center}
    \includegraphics[width=0.44\textwidth, trim=0cm 0cm 0cm 0cm, clip]{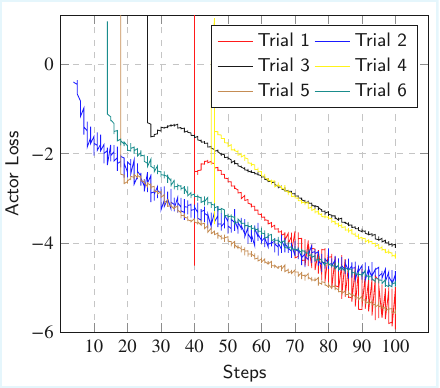}
\end{center}
        \caption{Loss variation across training steps for different trials of the actor network.}
    \label{fig:actor_loss_plot} 
\end{figure}

\begin{figure}[!t]
\centering
\begin{center}
    \includegraphics[width=0.46\textwidth, trim=0cm 0cm 0cm 0cm, clip]{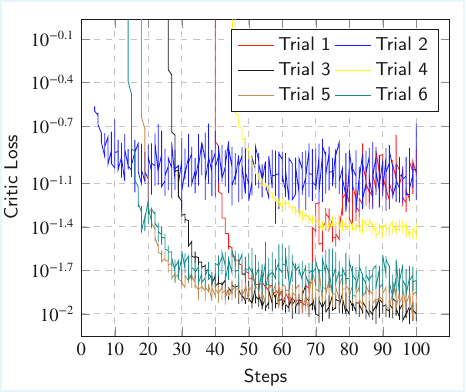}
\end{center}
    \caption{Loss variation across training steps for different trials of the critic network.}
    \label{fig:critic_loss_plot} 
\end{figure}

The performance of the DDPG reinforcement learning framework was evaluated across multiple trials with varying hyperparameter configurations, as detailed in the random search strategy. The evolution of Actor Loss and Critic Loss over training steps for six selected trials is depicted in figures \ref{fig:actor_loss_plot} and \ref{fig:critic_loss_plot}, respectively.

From figure \ref{fig:actor_loss_plot}, it is evident that \textit{Trial 5} achieves the most significant reduction in Actor Loss, steadily converging to a value of approximately \(-6\) by the final steps. This performance highlights its ability to optimize the actor network effectively, even under complex vehicular environments. Similarly, figure \ref{fig:critic_loss_plot} reveals that \textit{Trial 5} exhibits one of the lowest Critic Loss values, with stable convergence and minimal variance, underscoring the robustness of its critic network training. Comparatively, other trials, such as \textit{Trial 1} and \textit{Trial 2}, show slower convergence and higher variability in both Actor and Critic Loss metrics, likely attributed to suboptimal hyperparameter configurations. The superior performance of Trial 5 can be attributed to its carefully balanced hyperparameter set, specifically a shared learning rate of 0.00226 for both the actor (\(\alpha_{\pi}\)) and critic (\(\alpha_Q\)) networks, a discount factor (\(\gamma\)) of 0.882, a batch size of 272, and a replay buffer size of 30,397. This configuration allowed \textit{Trial} 5 to effectively balance exploration and exploitation during training, achieving both stability and efficiency in resource allocation. Therefore, \textit{Trial 5} emerges as the best-performing model, demonstrating its suitability for deployment in vehicular networks due to its efficient convergence and robust loss minimization. This selection highlights the critical role of hyperparameter tuning in ensuring the success of reinforcement learning-based systems.
    

\section{Results and Discussion}
\label{sec:results}
In this section, we present the numerical results derived from our proposed approach. To evaluate the performance of the hybrid framework, we compare it with standalone metaheuristic AO and DDPG-based AC-RL model in terms of convergence behavior, SE, and fairness.

\begin{figure}[!t]
\centering
\begin{center}
    \includegraphics[width=0.44\textwidth, trim=0cm 0cm 0cm 0cm, clip]{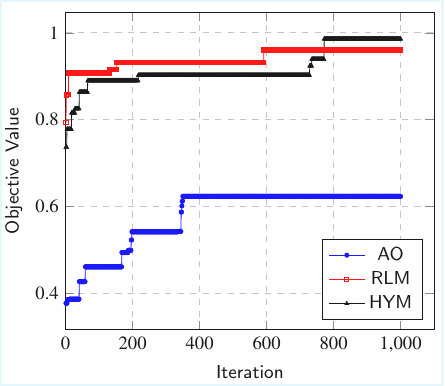}
\end{center}
\caption{Objective value convergence across AO, RLM, and HYM.}

\label{fig:objective_comparison}
\end{figure}

\begin{figure}[!t]
\centering
\begin{center}
    \includegraphics[width=0.44\textwidth, trim=0cm 0cm 0cm 0cm, clip]{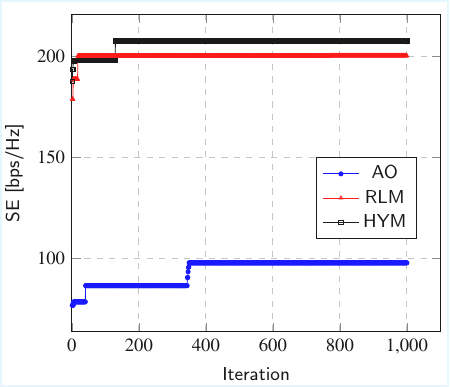}
\end{center}
\caption{Evolution of SE across AO, RLM, and HYM.}
\label{fig:throughput_comparison}
\end{figure}

Figure \ref{fig:objective_comparison} illustrates the convergence of the objective function for the three considered approaches: the AO, the reinforcement learning model (RLM), and the proposed hybrid RL-metaheuristic (HYM). As observed in figure \ref{fig:objective_comparison}, the convergence trends highlight the differences in optimization behavior among these approaches. The AO exhibits the slowest convergence, stabilizing at an objective value of approximately 0.6, as it lacks an adaptive mechanism for refining decisions based on prior outcomes, making it prone to local optima entrapment. In contrast, the RLM demonstrates a significantly higher convergence rate, reaching approximately 0.95 due to its iterative learning process, which progressively improves its decision-making policy. However, minor fluctuations in the optimization trajectory suggest occasional instability in maintaining the optimal solution. The proposed hybrid approach outperforms both standalone methods by achieving the fastest and most stable convergence, rapidly reaching an objective value close to 1.0. This superior performance results from integrating AO's global search capability with RLM’s adaptive learning mechanism, ensuring a balanced trade-off between exploration and exploitation.  
Beyond the convergence analysis, it is also crucial to examine the impact of each approach on system throughput, as efficient resource allocation should ultimately translate into improved network performance. Figure \ref{fig:throughput_comparison} presents the throughput evolution over iterations for AO, RLM, and HYM, providing further insights into their efficiency in optimizing CFmMIMO resource allocation. The observed trends confirm the disparities in performance between the three approaches. The AO exhibits the lowest throughput values, stabilizing at approximately 100 bps/Hz, which aligns with its slower convergence and limited adaptability. While RLM significantly improves throughput, reaching values close to 200 bps/Hz, minor fluctuations indicate occasional inefficiencies in decision-making. In contrast, the proposed hybrid approach achieves the highest and most stable throughput, exceeding 200 bps/Hz, confirming its superior ability to allocate frequency resources efficiently. The combination of global search from AO and adaptive learning from RLM enables HYM to optimize throughput effectively while maintaining robustness against interference and frequency variations. 

\begin{figure}[!t]
\centering
\begin{center}
    \includegraphics[width=0.458\textwidth, trim=0cm 0cm 0cm 0cm, clip]{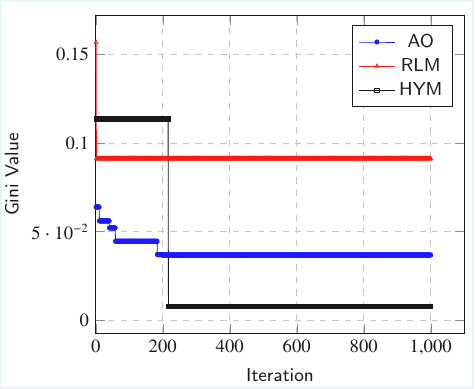}
\end{center}
\caption{Fairness comparison using Gini index for AO, RLM, and HYM over iterations.}
\label{fig:gini_comparison}
\end{figure}

While throughput is a key performance metric, fairness in resource allocation is equally important in CFmMIMO systems, particularly in multi-UE scenarios. Figure \ref{fig:gini_comparison} presents the Gini index evolution over iterations for the three considered approaches, providing insights into their fairness performance.

The Gini index is a well-established metric for fairness evaluation, where lower values indicate a more equitable distribution of resources among UEs. The results reveal notable differences among the three methods. The AO achieves a low Gini index, converging to approximately 0.05, suggesting that its resource allocation is more uniform. However, this comes at the cost of lower overall throughput, as previously discussed. The RLM stabilizes at a higher Gini index of around 0.1, indicating a slight imbalance in resource distribution, likely due to its learning-based prioritization of SE. Meanwhile, the proposed HYM approach achieves the best trade-off between fairness and performance, initially fluctuating but ultimately stabilizing at a Gini index of approximately 0.02. This behavior reflects its ability to balance UE fairness while maintaining high throughput, reinforcing its effectiveness as a resource allocation strategy.

\begin{figure}[!t]
\centering
\begin{center}
    \includegraphics[width=0.44\textwidth, trim=0cm 0cm 0cm 0cm, clip]{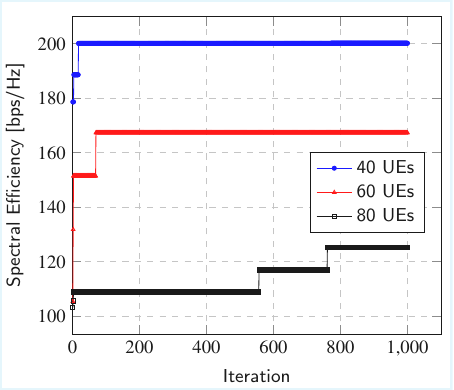}
\end{center}
\caption{SE comparison across 40, 60, and 80 UEs.}
\label{fig:SE_vs_UE}
\end{figure}

To further evaluate the scalability of the proposed hybrid approach in denser network conditions, we conducted an experiment where the number of resource blocks was fixed at 40, while the number of UEs was increased. Figure \ref{fig:SE_vs_UE} illustrates the SE performance for different UE densities, specifically for 40, 60, and 80 UEs. As observed, SE decreases as network congestion increases due to the heightened interference and competition for limited resources. When serving 40 UEs, the hybrid approach maintains a SE of approximately 200 bps/Hz, demonstrating its effectiveness in moderate UE density scenarios. With 60 UEs, SE declines to around 170 bps/Hz, reflecting the increased demand for the fixed resource blocks. In the most congested case with 80 UEs, SE further drops to approximately 110 bps/Hz, indicating a saturation point where resource limitations significantly constrain overall network performance.

Despite the expected degradation in SE with increased UE density, the proposed hybrid approach continues to ensure efficient resource allocation and system stability.

\begin{figure}[!t]
\centering
\begin{center}
    \includegraphics[width=0.44\textwidth, trim=0cm 0cm 0cm 0cm, clip]{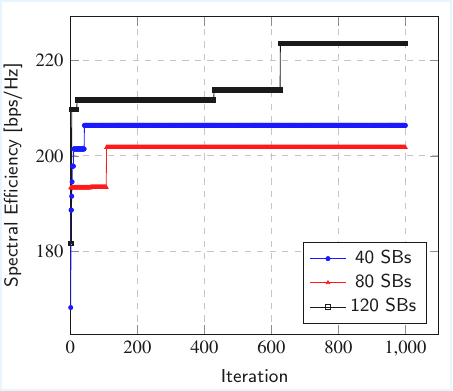}
\end{center}
\caption{SE comparison across 40, 80, and 120 SBs distributions.}
\label{fig:SB_comparison}
\end{figure}

To further investigate the impact of resource distribution on system performance, we analyze SE under varying numbers of subbands (SBs). While the previous analysis examined the effect of increasing UE density on SE, this section evaluates how allocating different SB quantities influences performance. As shown in Figure \ref{fig:SB_comparison}, increasing the number of SBs enhances SE, demonstrating the importance of resource availability in CF-mMIMO systems. When only 40 SBs are available, SE stabilizes at approximately 200 bps/Hz, aligning with the findings from the prior UE density analysis. Expanding the SB allocation to 80 SBs slightly reduces SE, converging around 195 bps/Hz due to increased interference and resource contention. However, with 120 SBs, SE reaches approximately 225 bps/Hz, showcasing the positive effect of a higher number of SBs on network efficiency. These results confirm that while a higher UE density negatively impacts SE due to increased competition, expanding SB allocation mitigates this degradation by optimizing spectral resource distribution. This highlights the adaptability of the proposed hybrid framework in dynamically managing resources to maintain robust performance across varying network conditions.

\section{Conclusions and Future Work }
\label{sec:conclusion}

In this work, we proposed a hybrid RL-metaheuristic framework for efficient resource allocation in CF-mMIMO systems, integrating the advantages of both deep RL and metaheuristic optimization. The proposed approach was evaluated against standalone AO and an AC-RL model in terms of convergence behavior, SE, and fairness. The results demonstrated that the hybrid framework achieved superior convergence speed and stability, reaching an objective value close to 1.0 while mitigating the risk of local optima entrapment. Additionally, the proposed approach consistently outperformed the baseline methods in SE performance, achieving over 200 bps/Hz under standard network conditions, while also ensuring fairness in resource allocation with a lower Gini index.

Beyond individual performance metrics, the study examined the impact of varying UE density and SB availability on system efficiency. The findings revealed that increasing UE density led to SE degradation due to heightened competition, while expanding SB allocation improved SE, confirming the framework’s adaptability in complex network conditions. The hybrid approach successfully balanced the trade-offs between global exploration and adaptive learning, making it a robust solution for CF-mMIMO resource allocation.

Future work will focus on extending the framework to multi-agent RL scenarios to further enhance distributed decision-making capabilities. Additionally, investigating the impact of real-time channel variations and incorporating energy efficiency constraints will provide deeper insights into the practical deployment of the proposed model in next-generation wireless networks.

\bibliographystyle{elsarticle-num} 

\bibliography{refs}


\bio{}
\textbf{Selina Cheggour} received an M.S. degree in Electronics and Telecommunications Systems from the University of Nice in 2022 and is currently pursuing her Ph.D. at the University of Lille. Her research focuses on 6G mobile networks, with an emphasis on antenna design, signal processing, and cell-free massive MIMO systems. She employs an AI-driven approach for resource allocation applications in vehicular networks for next-generation mobile communication systems.

\noindent \textbf{Valeria Loscri} (M'03, SM'17) is a research director at Inria Lille (France). She joined Inria in October 2013. From December 2006 to September 2013, she was a Research Fellow in the TITAN Lab of the University of Calabria, Italy. She received her M.Sc. and Ph.D. degrees in Computer Science in 2003 and 2007, respectively, from the University of Calabria, and her HDR (Habilitation à diriger des recherches) in 2018 from Université de Lille (France). Her research interests focus on emerging technologies for new communication paradigms such as Visible Light Communication (VLC), mmWave, cybersecurity in wireless networks, and the cooperation and coexistence of wireless heterogeneous devices. She is involved in several European projects, including Horizon Europe MLSysOps, H2020 CyberSANE, and FP7 EU project VITAL, as well as national projects such as ASTRID DEPOSIA and ANR NEMIoT.

In 2021, she was nominated as one of the Women Stars in Computer Networking and Communications by the IEEE Communication Society. In 2024, she was shortlisted as a Cyber Researcher by the European Women Cyber Day (ECWD). Since 2023, she has been the Action Chair and Scientific Holder of the BEING-WISE COST Action (\href{https://beingwise.eu/}{https://beingwise.eu/}).

Recently, she was recognized as one of the 100 brilliant women in 6G (\href{https://medium.com/@womenin6g/announcing-the-100-brilliant-and-inspiring-women-in-6g-list-for-2025-7115d8c8a9ce}{https://medium.com/@womenin6g/announcing-the-100-brilliant-and-inspiring-women-in-6g-list-for-2025-7115d8c8a9ce}).

She served as the European Partnership Scientific Delegate for Inria Lille from October 2016 to September 2019. She was also a member of the Technology Transfer Committee at Inria Lille (October 2016 – September 2018). 

Currently, she is the Chair of the Expert Panel for Fundamental Research W\&T5 Computer Science \& Information Technology FWO. Since 2019, she has been serving as the Scientific International Delegate for Inria Lille.
\endbio

\end{document}